\documentclass[preprint,5p,number,sort&compress]{elsarticle}

\usepackage{amssymb}
\usepackage{tikz}
\usepackage{pgfplots}
\usepgfplotslibrary{groupplots}
\usepackage{amsmath}
\usepackage{color}
\def\wt#1{\widetilde{#1}}
\def\({\left(\frac{}{}}
\def\){\frac{}{}\right)}
\def\[{\left[\frac{}{}}                                
\def\]{\frac{}{}\right]}
\def\ZZ{Ziolkowski }
\def\sqrr{ \sqrt{ k^2-(\frac{\omega}{c}  )^2 }  }
\def\sqr{ \sqrt{ (\frac{\omega}{c}  )^2 - k^2 }  }
\def\W{\ell_0}
\def\GAM{ \Gamma }
\def\EE{ \widehat{ \cal E  }}

\def\FF{{ \cal F }}

\def\CE{\mbox{\tiny CE}}
\def\CM{\mbox{\tiny CM}}
\def\params{$(\Lambda, \Psi_1,\Psi_2, \Phi,\Xi  )\,$}

\def\J{\mathcal{J}}
\def\BLP{Bopp-Land\'{e}-Podolsky }

\def\POWERPULSEPLOT{
\begin{figure}[!ht]
		\centering
		\includegraphics{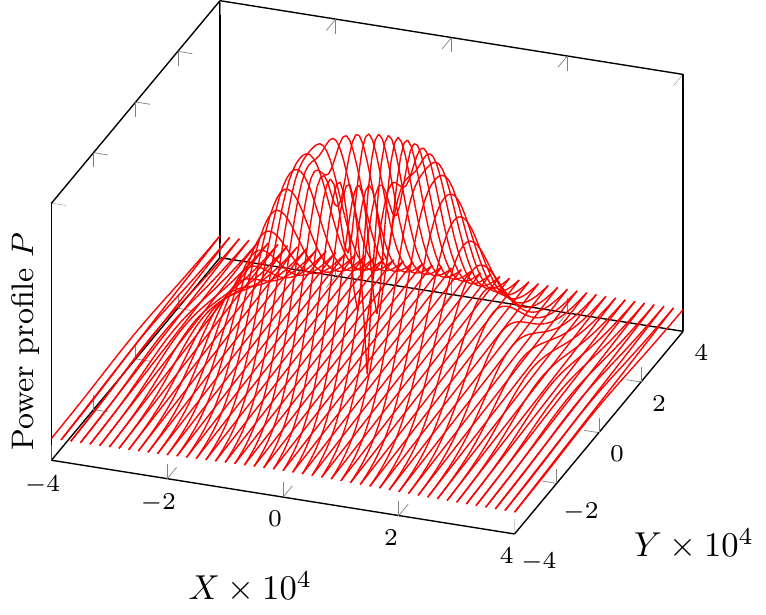}
	\caption{Power profile $P$  of the $(\text{CM},1)$ laser pulse at $Z=0,\,T=0$  with parameters $\{\Lambda=600, \Psi_{1}=1, \Psi_{2}=1000, \Phi=0.001, \Xi=1\}$}
	\label{LPPD}
\end{figure} 
}

\def\LASERSPACECURVES{
\begin{figure*}[!ht]
		\centering
		\includegraphics{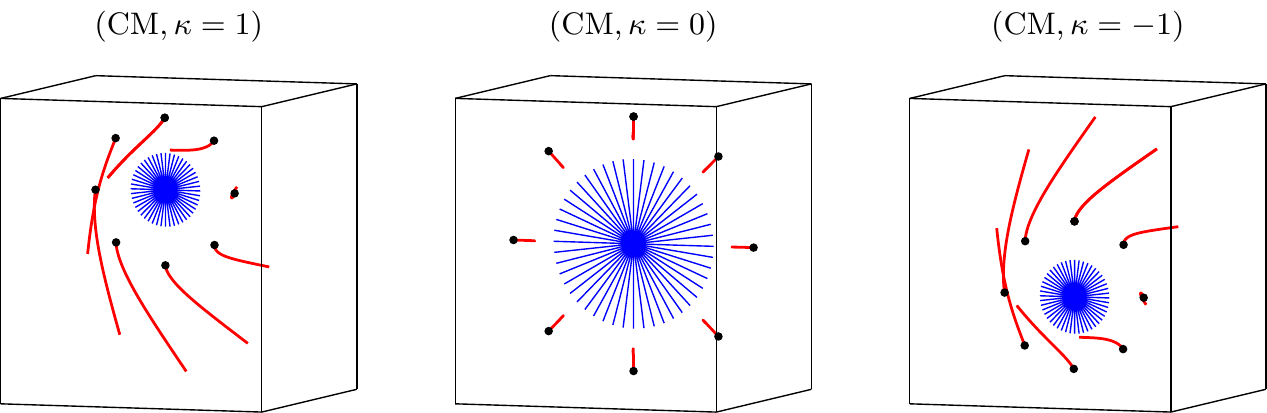}
	\caption{Three-dimensional spacecurves for particles subject to an incident $(\text{CM},1)$ laser pulse (left), $(\text{CM},0)$ laser pulse (centre) and $(\text{CM},-1)$ laser pulse (right) with parameters $\{\Lambda=600, \Psi_{1}=1, \Psi_{2}=1000, \Phi=0.001, \Xi=1\}$. Each particle has initial velocity $\{\dot{R}(0)=0, \dot{\theta}(0)=0, \dot{Z}(0)=\frac{1}{200}\}$. The shaded  circular disc region indicates the initial  spot size ($R=10000$ for $(\text{CM},\pm 1)$ laser pulses and $R=20000$ for a $(\text{CM},0)$ laser pulse) relative to the black markers on the spacecurves that denote the initial positions of the charged test particles.}
	\label{DTP}
\end{figure*} 
}

\def\LASERSKE{
\begin{figure*}[htbp]
		\centering
		\includegraphics{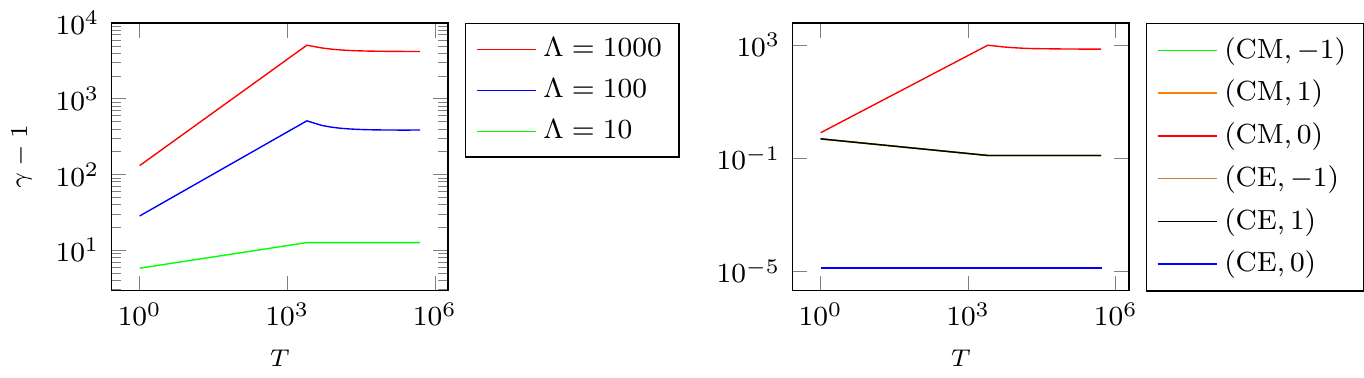}
	\caption{Specific kinetic energy transfer to any charged test particle. On the left, the (CM,$1$) pulse has parameters $\{\Psi_{1}=1, \Psi_{2}=1000, \Phi=0.001, \Xi=1\}$. On the right, relative values of $\gamma-1$ are displayed for various $(s,\kappa)$ pulses with parameters $\{\Lambda=1, \Psi_{1}=1, \Psi_{2}=1000, \Phi=0.001, \Xi=1\}$. The differences between the energy transfers for some pulses appear indistinguishable relative to others owing to the logarithmic scales employed. In all cases, the charged particle has initial position $\{R(0)=1, \theta(0)=\frac{\pi}{2}, Z(0)=1\}$ and initial velocity $\{\dot{R}(0)=0, \dot{Z}(0)=\frac{1}{200}, \dot{\theta}(0)=0\}$.}
	\label{SKE}
\end{figure*} 
}

\def\LASERTABLE{
	\begin{table*}[b]
	\centering
		\begin{tabular}{|c|c|c|c|c|c|c|c|c|}
			\hline
			 $s$ & ${\;\kappa\,\;}$ & ${\mathcal{E}}$ (J) & $\Lambda$ & ${z_{rg}}$ (m)\, & ${t_{0}}$ (ps) & ${z_{w}}$ (m) &  ${r_{s}(0)}$ (m) & $I$ (W\,$\text{cm}^{-2}$)\\
			\hline
			\hline
			CE & 0 & 8.07 & 40000 & 1.44 & 3 & $ 8.99\times 10^{-4}$ & 0.018 & $ 2.646 \times 10^{12}$ \\
			\hline
			CE & $\pm 1$ & 7.69 & 1 & 2.43 & 3 & $ 8.99 \times 10^{-4}$ & 0.009 & $ 1.009 \times 10^{12}$ \\
			\hline \hline
			CM & 0 & 8.41 & $\frac{1}{20}$ & 0.899 & 2 & $ 6.00\times 10^{-4}$ & 0.012 & $ 9.304 \times 10^{11}$ \\
			\hline
			CM & $\pm 1$ & 7.69 & 1 & 1.80 & 3 & $ 8.99 \times 10^{-4}$ & 0.009 & $ 1.009 \times 10^{12}$ \\
			\hline\hline
			CM & $\pm 1$ & 76903 & 100 & 1.80 & 3 & $ 8.99\times 10^{-4}$ & 0.009 & $ 1.009\times 10^{16}$ \\
			\hline
		\end{tabular}
		\caption{Table showing MKS laser characteristics for various $(s,\kappa)$ laser pulse configurations with parameters $\{\Psi_{1}=1, \Psi_{2}=1000, \Phi=0.001, \Xi=1\}$.}
		\label{MKSLPC}
	\end{table*}
}

\def\LASERTABLE{
	\begin{table}[b]
	\renewcommand{\arraystretch}{1.06}
	\centering
	\small
		\begin{tabular}{|c||c|c||c|c||c|}
			\hline
			 $s$ & CE & CE & CM & CM & CM \\
			\hline
			${\;\kappa\,\;}$  & 0 & $\pm 1$ & 0 & $\pm 1$ & $\pm 1$ \\
			\hline
			${\mathcal{E}}$ (J) & 8.07 & 7.69 & 8.41 & 7.69 & 76903 \\
			\hline 
			$\Lambda$ & 40000 & 1 & 0.05 & 1 & 100 \\
			\hline
			${z_{rg}}$ {\footnotesize (m) }\, & 1.44 & 2.43 & 0.899 & 1.80 & 1.80 \\
			\hline
			${t_{0}}$ {\footnotesize(ps)} & 3 & 3 & 2 & 3 & 3 \\
			\hline
			${z_{w}}$ {\footnotesize(mm) } & 0.899 & 0.899 & 0.600 & 0.899 & 0.899 \\
			\hline
			${r_{s}(0)}$ {\footnotesize(m)} & 0.018 & 0.009 & 0.012 & 0.009 & 0.009 \\
			\hline
			$I$ {\footnotesize (TW\,$\text{cm}\!{}^{-2}$)} & 2.646 & 1.009 & 0.930 & 1.009 & 10090 \\
			\hline
		\end{tabular}
		\caption{Table showing MKS laser characteristics for various $(s,\kappa)$ laser pulse configurations with parameters $\{\Psi_{1}=1, \Psi_{2}=1000, \Phi=0.001, \Xi=1\}$.}
		\label{MKSLPC}
	\end{table}
}

\begin{document}

\title{Classical Dynamics of Free Electromagnetic Laser Pulses}
\author[IMS]{S. Goto}
\ead{sgoto@ims.ac.jp}
\author[LU]{R. W. Tucker}
\ead{r.tucker@lancaster.ac.uk}
\author[BU]{T. J. Walton}
\ead{t.walton@bolton.ac.uk}
\address[IMS]{Institute for Molecular Science, 38 Nishigo-Naka, Myodaiji, Okazaki 444-8585, Japan}
\address[LU]{Department of Physics, University of Lancaster and Cockcroft Institute, Daresbury Laboratory, Warrington, UK}
\address[BU]{Department of Mathematics, University of Bolton, Deane Campus, Bolton, UK}
\begin{abstract}
	We discuss a class of exact finite energy solutions to the vacuum source-free Maxwell field equations as models for multi- and single cycle laser pulses in classical interaction with relativistic charged test particles. These solutions are classified in terms of their \textit{chiral} content  based on their influence on particular charge configurations in space. Such solutions offer a computationally efficient parameterization of compact laser pulses used in laser-matter simulations and provide a potential means for experimentally bounding the fundamental length scale in the generalized electrodynamics of Bopp, Land\'{e}  and Podolsky.
\end{abstract}
\maketitle
\setlength{\mathindent}{0.8cm}

Advances in laser technology have made possible the exploration of physical processes on unprecedented temporal and spatial scales. They have also opened up new possibilities for accelerating charged particles using laser-matter interactions. Multi- and single cycle high intensity ($10^{10}- 10^{15}$  Watts/cm${}^2$) laser pulses can be produced using Q-switching or mode-locking techniques \cite{brabec2000intense}. Such pulses can accelerate charged particles such as electrons  to relativistic speeds where radiation reaction and quantum effects may influence their dynamics. Lower intensity pulses have also been used as diagnostic tools for exploring the structure of plasmas in various states \cite{buck2011real,matlis2006snapshots}. In order to interpret experimental data involving classical laser interactions with both charged and neutral matter, theoretical models \cite{mackenroth2010determining,tsung2002generation,ziolkowski1991collective,ziolkowski2006metamaterial} rely crucially on parameterizations of the electromagnetic fields in laser pulses, particularly in situations where traditional formulations using monochromatic or paraxial-beam approximations have limitations \cite{liu2011exact,harvey2011symmetry,terranova2014particle}.

In this Letter we discuss a viable methodology for parameterizing a particular class of propagating solutions to the source free {\it classical } Maxwell equations in vacuo that offers an efficient  means to explore the  classical effects of compact laser pulses on free electrons in dynamical regimes where quantum effects are absent. The parameterization is based on a remarkable class of explicit solutions of the scalar wave equation found by \ZZ \cite{ziolkowski1985exact,ziolkowski1989localized,shaarawi1989localized,donnelly1992method,donnelly1993designing} following pioneering work by Brittingham \cite{brittingham1983focus}. Such solutions can be used to construct classical Maxwell solutions with bounded total electromagnetic energy and fields bounded in all three spatial directions. With simple analytic structures their diffractive properties can be readily determined together with the behaviour of charged particle-pulse interactions  over a broad parameter range without recourse to expensive numerical computation. Finally, we argue that such parameterizations can be used to find compact finite energy solutions to other linear wave equations. This is illustrated by showing that the generalized theory of Bopp \cite{bopp1940lineare}, Land\'{e} \cite{lande1941finite} and Podolsky \cite{podolsky1942generalized} admits such particular solutions that  reduce to the Maxwell solutions when a fundamental length parameter in their theory tends to zero. Compact laser pulses in this theory might be used to explore properties of the theory by searching experimentally for bounds on this parameter.

If a complex scalar field $\alpha $ satisfies $\Box\,\alpha=0$  on spacetime and $ \Pi_{\mu\nu}$ is any covariantly constant (degree 2) anti-symmetric tensor field on spacetime (i.e. $\Pi_{\mu\nu ;\delta }=0$)
for all $\mu,\nu,\delta =0,1,2,3  $, then the complex tensor field $F_{\mu\nu}=\partial_{\mu} A_{\nu} -  \partial_{\nu} A_{\mu} $ satisfies the source free Maxwell equations in vacuo with:
\begin{eqnarray}\label{ALP}
	A_{\nu}=\partial_{\gamma}\left (  \alpha \,\,\Pi_{\mu\beta  }  \right)  \, \epsilon^{\gamma\mu\beta   }{}_{\nu}\, \sqrt{ \vert  g \vert }
\end{eqnarray}
where $\vert g \vert$ is the determinant of the spacetime metric and $\epsilon^{\gamma\mu\beta}{}_{\nu}\,$ denotes the Levi-Civita alternating symbol. In the following $g$ refers to the Minkowski metric tensor field, in which case the components $\Pi_{ \mu\nu }$ can be used to encode three independent Hertz vector fields and their duals\footnote{In the language of differential forms on Minkowski spacetime $A=\star d(\alpha \Pi),\,\, F=dA$ where $d\star d\,\alpha=0$, the 2-form $\Pi$ satisfies $\nabla \Pi =0$ and $\star$ denotes the Hodge map associated with $g$.}.

\POWERPULSEPLOT

General solutions to  $\Box \,\alpha =0$ can be constructed by Fourier analysis. In cylindrical polar Minkowski coordinates $\{t,r,z,\theta\}$, axially symmetric solutions propagating along the $z$-axis have, for $z\geq 0$, the double integral representation $\alpha(t,r,z)=\int_{-\infty}^{\infty} \, d\omega \, e^ {-i \omega t} \wt{\alpha}(\omega,r,z)$ where:
\setlength{\mathindent}{0.3cm}
\begin{eqnarray*}
	\wt{\alpha}(\omega,r,z) &=& \int_0^{ \frac{\omega}{c} }  k \, f_\omega(k) J_0(k r) e^{ \pm i z \sqr  }   \, dk   \\[0.2cm]
				&& + \int_{ \frac{\omega}{c} } ^ \infty k f_\omega(k) J_0(k r) e^{ - z \sqrr  }   \, dk      
\end{eqnarray*}
in terms of the zero order Bessel function and the speed of light in vacuo $c$. 

\LASERSPACECURVES

Conditions on the Fourier amplitudes $f_\omega(k)$ can be given so that the Hertz procedure above gives rise to real singularity free electromagnetic fields with finite total electromagnetic energy. A particularly simple class of pulses that can be generated in this way follows from the complex axi-symmetric scalar solution:
\begin{eqnarray}\label{alp}
   \alpha(t,r,z) &=& \frac{\W^2}{r^2 + ( \psi_1 + i(z-ct)  )\,(  \psi_2-i( z+ct )  ) }
\end{eqnarray}
where $\W, \psi_1,\psi_2$ are arbitrary real positive definite parameters with physical dimensions of length. The relative sizes of $\psi_1$ and $\psi_2$  determine  both the direction of propagation  along the $z-$axis of the dominant maximum of the pulse profile and the  number of  spatial cycles in its peak magnitude. When $\psi_{1}\gg\psi_{2}$, the dominant maximum propagates along the $z-$axis to the right. The parameter $\ell_0$ determines the magnitude of such a maximum. The structure of such solutions has been extensively studied in \cite{fedotov2007exact,feng1999spatiotemporal} in conjunction with particular choices of $\Pi_{\mu\nu}$ together with generalizations discussed in \cite{borzdov2002designing,hernandez2007localized}.

In general the six anti-symmetric tensors with components $\delta^\mu_{ [\gamma } \delta^\nu_{\sigma ]}$ in a Minkowski Cartesian coordinate system are covariantly constant and can be used to construct a complex eigen-basis of antisymmetric {\it chiral} tensors $ \Pi^{s,\,\kappa }$, with  $s \in \{\text{CE},\text{CM}\} $ and $\kappa \in \{ -1,0,1 \}$,  satisfying
\setlength{\mathindent}{0.8cm}
\begin{eqnarray} \label{chiral}
	{\cal O}_z  \, \Pi^{s,\,\kappa } = \kappa\, \Pi^{s,\,\kappa }
\end{eqnarray}
where the operator ${\cal O}_z$ represents  $\theta$ rotations about the $z-$axis generated by $-i \partial_\theta $ on tensors\footnote{In terms of the Lie derivative, ${\cal O}_z=-i {\cal L }_{\partial_{\theta} }$  and
$\Pi^{ \CE,\pm 1 }=d(x\pm i y)\wedge dt,\,\, \Pi^{ \CE,0 }=dz\wedge dt, \Pi^{ \CM,\kappa } =  \star \Pi^{ \CE,\kappa }$ where $x=r \cos(\theta), y=r \sin(\theta)$}. These in turn can be used to construct a {\it complex basis} of chiral eigen-Maxwell tensor fields $F^{s,\,\kappa}$. The index $s$ indicates that the CE (CM) chiral family contain electric (magnetic) fields that are orthogonal to the $z-$axis when $\kappa=0$. The chiral eigen-fields $F^{s,\,0}$ inherit the axial symmetry of $\alpha(t,r,z)$ while those with $\kappa=\pm 1$ do not. The directions of electric and magnetic fields in any of these  Maxwell solutions depend on their location in the pulse and the concept of a pulse polarisation is not strictly applicable. The chiral content as defined here can be used in its place. Non-chiral pulse configurations can be constructed by superposition $\sum_s\, \sum_\kappa F^{s,\,\kappa}  \, {\cal C}^{s,\,\kappa}$ with arbitrary complex coefficients ${\cal C}^{s,\,\kappa}$.

The energy, linear and angular momentum of the pulse in vacuo can be calculated from the components $T_{\mu\nu}$ of the Maxwell stress-energy  tensor $T_{\mu\nu}=-\frac{1}{4} g_{\mu\nu} \FF^{\alpha\beta}\FF_{\alpha\beta} - \FF_{\mu\alpha} \FF^{\alpha}{}_{\nu} $
where $\FF_{\mu\nu}= \text{Re}(F_{\mu\nu})$. If $\mathbf{e}$ and $\mathbf{b}$ denote time-dependent \textit{real} electric and magnetic 3-vector fields associated with any pulse solution,  its total electromagnetic energy  $\J$, for a fixed set of parameters  and any $z$,  is calculated from
\begin{eqnarray}\label{P}
	\J &=& \frac{1}{\mu_0} \int_{-\infty}^{\infty} \, dt \, \int_{S}  (  { \mathbf e} \times {\mathbf b}   ) \cdot d{\mathbf S}
\end{eqnarray}
where $S$ can be any plane with constant $z=z_0>0$. For {\it spatially compact} pulse fields in vacuo this coincides with the total pulse electromagnetic energy
\setlength{\mathindent}{0.3cm}
\begin{eqnarray}\label{ENERGY}
	{\cal E} &\!\!=\!\!& \int_{\cal V}  \rho\, d{\cal V} \,=\, \int_{-\infty}^{\infty}\!\! dz\,\int_0^{ 2 \pi} \!\!d\theta\, \int_{ 0} ^\infty\!\! r dr\;\rho(t,r,z,\theta)
\end{eqnarray}
where $\rho\equiv \frac{1}{2}\,\left( \epsilon_0\, {\mathbf e} \cdot {\mathbf e}  + \frac{ {\mathbf b} \cdot {\mathbf b}}{\mu_0 } \right)$ is integrated over all space ${\cal V}$. This follows since $\nabla\cdot({ \mathbf e} \times { \mathbf b} ) = -\mu_0\,\partial_t\,\rho $. To correlate $\J$ with other laser pulse properties and the choice of parameters, we bring the pulse into classical interaction with one or more charged point particles. The world-line of a single particle, parameterized in arbitrary coordinate as $x^\mu= \xi^\mu(\tau)$ with a parameter $\tau$, is taken as a solution of the coupled non-linear differential equations
\setlength{\mathindent}{0.8cm}
\begin{eqnarray} \label{EOM}
	{ { \cal A}_\mu  (\tau) }= \frac{q}{m_0 c^2}\FF_{\mu\nu}( \frac{}{} \!\xi(\tau) ) \, V^\nu(\tau)
\end{eqnarray}
in terms of the particle charge $q$ and rest mass $m_0$, for some initial conditions $\xi(0), V(0)$, where the particle 4-velocity satisfies $V^\nu\, V_\nu=-1$ and its 4-acceleration is expressed in terms of the Christoffel symbols $\Gamma^{\delta}{}^\beta{}_{\mu}$ as  ${  \cal A}_\mu = \partial_\tau \, {V_\mu}(\tau) +  V_{\delta} (\tau)\,V_{\beta}(\tau)\, \Gamma^{\delta}{}^\beta{}_{\mu}( \xi( \tau) )$.  In the following, radiation reaction and inter-particle forces are assumed negligible. From the solution $\xi(\tau)$ one can determine the increase (or decrease) in  the relativistic kinetic energy transferred from the electromagnetic pulse to any  particle and the nature of its trajectory in the laboratory frame.  This information can then be used to correlate the dynamical properties of the interaction with the laser pulse properties fixed by the parameters. To facilitate this exercise, it proves important to reduce the above equations of motion to dimensionless form and fix the physical dimensions of the fields involved. The Minkowski metric tensor field $g=g_{ \mu\nu} dx^\mu\, dx^\nu$
(with $g_{ \mu\nu }=\mathrm{ diag } (-1,1,1,1)$)  in inertial coordinates $x^0=c t,x^1=x,x^2=y,x^3=z$) has MKS physical dimensions $[L]^2$. The MKS dimension of electromagnetic quantities follows by assigning to $ \epsilon_0 F_{\mu\nu}\, dx^{[\mu} dx^{\nu]}$ in any coordinate system the  physical of electric charge dimension.  Furthermore, in terms of Minkowski polar coordinates $\{ t,r,z,\theta \}$, introduce (for ease of visualization) the dimensionless coordinates $\{R=\frac{r}{\Phi\W}, T=\frac{c t}{\W}, Z=\frac{z}{\Xi\W}\}$ and dimensionless parameters $\Lambda, \Psi_j=\frac{\psi_j }{\W}$ ($j=1,2$) where ${[\Psi_j]= [\Phi]=[\Xi]=1},{[\W]=[L]}$. Then with the dimensionless complex scalar field  $\widehat{\alpha }(T,R,Z)=\alpha(t,r,z)$  and greek indices ranging over $\{ T, R, Z,\theta \}  $ with $ \epsilon^{ T,R,Z,\theta  } = 1$ ,  we write
\begin{eqnarray} \label{dimless}
	A_\delta &=& \frac{m_0 c^2 \W^3\, \Lambda}{q} \,\,\partial_\gamma\left(\widehat{ \alpha} \, \widehat{ \Pi}_{\mu\beta }\right) \epsilon^{\gamma\mu\beta}{}_\delta \sqrt{ \vert g \vert  }
\end{eqnarray}
for a choice of dimensionless covariantly constant tensor $\widehat{\Pi}_{\mu\beta }$ so that $ [{\epsilon_0}A_\mu\, dx^\mu]$  has the physical dimension of electric charge  and
\begin{eqnarray*}
	\J 		 &=& \int_{-\infty}^\infty\,dT\,\int_0^\infty\,d R\,\int_0^{ 2\pi}\,d\theta\,P(T,R,Z,\theta) \\[0.2cm]
	{\cal E} &=& \int_{-\infty}^\infty \,dZ\,\hat{ \cal E}(T,Z).
\end{eqnarray*}
The parameter $\Lambda$ controls the strength of all electric and magnetic fields in $F_{\beta\delta}$ for fixed values of the parameters $\Psi_1,\Psi_2,\Phi,\Xi$ and the overall scale $\W$ will be fixed in terms of the total electromagnetic energy of the pulse. For a choice of such  parameters the real fields $ {\mathbf e }$ and ${\mathbf b } $ enable one to calculate a numerical value $\GAM$ such that $\J=\W\GAM$. The diffraction of the pulse peak along the $z-$axis can be used to define a pulse range relative to the maximum of the pulse peak at $z=0$. To this end, the density $\EE( T,Z )$ defines the dimensionless range $Z_{rg}$ by ${\EE(0,0  ) }/{\EE( T_1,Z_{rg} )} = {2}$, where the peak at $Z=Z_{rg} > 0$ and $T=T_1 >0 $ is half the height of the peak at $Z=0, T=0$. If during the interval $[0,T_1]$ the pulse propagates with negligible deformation in $Z$, one may estimate its width $Z_w$ at half height and the dimensionless pulse axial speed $\beta=Z_{rg}/T_1$. This yields the dimensionless pulse duration or temporal width $T_0={Z_w}/{\beta}$. From these dimensionless values one deduces the pulse MKS characteristics in terms of $\W$ and hence $\J$. If the picosecond is used as a unit of time, the pulse duration becomes $t_0= \W T_0/c= \W Z_w/(\beta c)=N\, 10^{-12}$ sec for some value $N$ and hence $\W=(c \beta N/Z_w)\, 10^{-12}$ metres,  $\J= (\GAM {\beta }c N/Z_w) \, 10^{-12}$  Joules,  $z_{rg}= \W\Xi Z_{rg}=(\Xi{\beta} c N Z_{rg}/Z_w)\,10^{-12}  $ metres  and $z_w=\Xi c \beta n\,10^{-12} $ metres.
A dimensionless spot-size of the pulse at $ Z=Z_0 >0 ,\, T=Z_0/\beta $ is  then determined by  the behaviour of $ P(R,Z_0/\beta, Z_0,\theta)$.  At each value of $Z_0$ this function of $R$ and $\theta$ has a clearly defined principal maximum. If one associates a circle of dimensionless radius $R_s(Z_0)$ with such a maximum locus it can be used to define a spot-size with radius $r_s(z_0)\!=\!\W\Phi R_s(Z_0)\!=\! (c \beta N \Phi R_s(Z_0)/Z_w)10^{-12}$ metres at $z=z_0$.

\LASERSKE

Figure~\ref{LPPD} displays a clearly pronounced principle maximum in the power density  profile $P$ as a function of $X=R\cos(\theta)$ and $Y=R\sin(\theta)$ at $Z=0,T=0$ for a specific choice of the parameters \params. The same parameter set is used to numerically solve (\ref{EOM}) for a collection of  trajectories for charged particles, each arranged initially around the circumference of a circle in a plane orthogonal to the propagation axis of incident CM type laser pulses with different chirality. The resulting space curves in 3-dimensions, displayed in figure~\ref{DTP}, clearly exhibit the different responses to CM pulses with distinct chirality values. The instantaneous {\it specific} relativistic kinetic energy of a particle with laboratory speed $v$ is $\gamma - 1$ in terms of the Lorentz factor $\gamma$ given by $\gamma^{-1}=\sqrt{1-\frac{v^2}{c^2}}$. \LASERTABLE 
In figure~\ref{SKE}, this quantity is displayed on the left for a charged particle accelerated by a fixed chirality (CM,$-1$) type pulse where the pulse energy is varied by changing $\Lambda$. On the right the energy transfer dependence on pulse chirality for both CE and CM type pulses with  {\it fixed laser energy} is displayed. We deduce that the momentum and angular momentum \cite{allen2003optical} in the propagation direction can transfer an impulsive force and torque respectively to charges lying in an orthogonal plane. More generally, the classical configurations of a high energy pulse labelled CE and CM can be distinguished by their interaction with different arrangements of charged matter.

Furthermore, by a suitable choice of parameters, (CE, $\kappa$) type modes can be constructed that yield the same physical properties ($\J,z_{rg}, z_w,\beta$) for all $\kappa$. Similarly the (CM,$\kappa$) type modes yield a $\kappa$ independent set with physical properties distinct from those determined by the (CE,$\kappa$) modes. The pulse  group speed magnitudes (as defined above) of all these configurations are determined numerically and are bounded above by the value $c$. To illustrate some of these statements, table~\ref{MKSLPC} summarizes the MKS laser pulse characteristics for a specific choice of $\{\Psi_1,\Psi_2, \Phi,\Xi\}\,$ and various values of $\Lambda$.
  
Motivated by the desire to ameliorate the divergences in perturbative QED a number of generalized theories of electromagnetism have been proposed. To date, there is little experimental evidence for testing their predicted departures from Maxwell's theory. However, with the increase in laser technology one may now be entering regimes that may discriminate between such theories. In particular, \BLP electrodynamics is linear in the electromagnetic fields, contains a fundamental length $\lambda_{0}$ and approaches Maxwell's theory when this length tends to zero. Unlike Maxwell's theory, it contains solutions describing static point charges  with \textit{finite} electromagnetic energy. \\

The classical \, source-free \, \BLP \,field \\ equations in vacuo for the complex electromagnetic field tensor $F_{\mu\nu}=\partial_{\mu}A_{\nu}-\partial_{\nu}A_{\mu}$ are:
\begin{eqnarray}\label{BLP}
	\partial^{\mu}F_{\mu\nu} - \lambda_{0}^{2}\,\Box\, \partial^{\mu}F_{\mu\nu} &=& 0.
\end{eqnarray}
Clearly, all classical source-free vacuum Maxwell solutions satisfy these equations. However, there exist additional solutions to (\ref{BLP}) that are \textit{not} source-free vacuum Maxwell solutions. For example, such additional solutions of (\ref{BLP}) follow from (\ref{ALP}) provided:
\begin{eqnarray}\label{BLPalpha}
	\Box\,\alpha + \lambda_{0}^{2}\,\Box^{2}\,\alpha &=& 0.
\end{eqnarray}
Furthermore, it follows that as $\lambda_{0}\rightarrow 0$, one recovers the class of Maxwell solutions discussed above. For non-zero $\lambda_{0}$, there exist particular non-Maxwellian solutions to (\ref{BLPalpha}) satisfying
\begin{eqnarray*}
	\Box\,\alpha + \frac{1}{\lambda_{0}^{2}} \alpha &=& 0.
\end{eqnarray*}
In Minkowski cylindrical coordinates $\{t,r,\theta,z\}$, a non-separable axi-symmetric compact laser solution is given for arbitrary real $\Lambda_0$ by:
\begin{eqnarray*}
	\alpha(t,r,z) &=& \frac{\Lambda_{0}^{2}\,K_{1}\( \frac{\zeta(r,t,z)}{\lambda_{0}} \)}{ \zeta(t,r,z) }
\end{eqnarray*}
where 
\setlength{\mathindent}{0.2cm}
\begin{eqnarray*}
	\zeta(t,r,z) &\equiv& \sqrt{ r^{2} + \[\psi_{1} + i(z-ct)\]\[\psi_{2} - i(z+ct)\]}
\end{eqnarray*}
in terms of arbitrary real positive-definite parameters $\psi_{1}$, $\psi_{2}$ that determine propagation along the positive $z$-axis and where $K_{1}(z)$ denotes the first order modified Bessel function of the second kind. The function $\zeta(t,r,z)$ is positive for all real $r,t,z$ and hence both the real and imaginary parts of $\alpha(t,r,z)$ are bounded.  \\

The interaction of the \BLP pulse with classical charged test particles follows from the divergence of the \BLP stress-energy-momentum tensor   \cite{Tucker_BP} . This in turn leads to the same equation of motion (\ref{EOM}) for the test particle trajectories \cite{Tucker_BP} and may offer an experimental means to discriminate between the Maxwell and \BLP descriptions of source-free electromagnetic fields in vacuo. \\

Further details of this approach using scalar field modulated Hertz potentials based on the construction (\ref{ALP}) for finding spatially compact solutions to other linear tensor and spinor wave equations will be presented elsewhere. \\

The authors are grateful to STFC (ST/G008248/1), EPSRC (EP/J018171/) and Dino Jaroszynski for support. They also thank Volker Perlick, Jonathan Gratus, David Burton and their other colleagues in the ALPHA-X project for useful discussions. Moreover, SG gratefully acknowledges support from Grant-in-Aid for Young Scientists \\ (Grant No. 25800181).\\

\bibliographystyle{unsrt}
\bibliography{Laser_PIPAMON}

\end{document}